# CTLformer: A Hybrid Denoising Model Combining Convolutional Layers and Self-Attention for Enhanced CT Image Reconstruction

**Zhiting Zheng*[1,4], Shuqi Wu[2,5], Wen Ding[3,6]**

[1] Independent Researcher, Los Angeles, CA, USA
[2] Amazon.com, Inc., New York, USA
[3] H. Milton Stewart School of Industrial and Systems Engineering, Georgia Institute of Technology, Atlanta, GA, USA

[4] gulingli0930@gmail.com

5 shuqiwu2023@u.northwestern.edu

6 wding5662@gmail.com

**Abstract.** Low-dose CT (LDCT) images are often accompanied by significant noise, which negatively impacts image quality and subsequent diagnostic accuracy. To address the challenges of multi-scale feature fusion and diverse noise distribution patterns in LDCT denoising, this paper introduces an innovative model, CTLformer, which combines convolutional structures with transformer architecture. Two key innovations are proposed: a multi-scale attention mechanism and a dynamic attention control mechanism. The multi-scale attention mechanism, implemented through the Token2Token mechanism and self-attention interaction modules, effectively captures both fine details and global structures at different scales, enhancing relevant features and suppressing noise. The dynamic attention control mechanism adapts the attention distribution based on the noise characteristics of the input image, focusing on high-noise regions while preserving details in low-noise areas, thereby enhancing robustness and improving denoising performance. Furthermore, CTLformer integrates convolutional layers for efficient feature extraction and uses overlapping inference to mitigate boundary artifacts, further strengthening its denoising capability. Experimental results on the 2016 National Institutes of Health AAPM Mayo Clinic LDCT Challenge dataset demonstrate that CTLformer significantly outperforms existing methods in both denoising performance and model efficiency, greatly improving the quality of LDCT images. The proposed CTLformer not only provides an efficient solution for LDCT denoising but also shows broad potential in medical image analysis, especially for clinical applications dealing with complex noise patterns.

## 1.Introduction

Computed Tomography (CT) is a high-precision clinical imaging tool that uses rotating X-ray beams or other penetrating radiation to scan the body. Detectors collect the signals, and reconstruction algorithms generate high-resolution cross-sectional images, which can be combined into 3D structures. CT offers better tissue resolution and faster scanning than traditional X-ray imaging, making it particularly useful in diagnosing neurological, thoracoabdominal diseases, trauma, and tumors. However, its widespread use raises concerns about radiation dose safety. Long-term exposure to high-dose ionizing radiation may increase cancer risk, especially for children and patients needing multiple scans. Recently, Low-Dose CT (LDCT) technology has gained attention. It reduces radiation dose but causes image noise and detail loss. Advanced algorithms like deep learning (CNN, Transformer, etc.) have been introduced in biomedical image processing to solve these issues, with image denoising, super-resolution reconstruction, and structure enhancement for LDCT images becoming hot research topics [1,2].

Low-Dose Computed Tomography (LDCT) has been developed as an alternative to reduce the X-ray dose. The radiation used in LDCT is significantly lower than that in traditional CT, approximately one-quarter of the dose used in standard CT, resulting in less radiation-induced harm to the body[3].

Traditional low-dose CT (LDCT) denoising methods often rely on physical models and prior knowledge through iterative algorithms. While effective in theory, these methods are computationally intensive and challenging to deploy in commercial CT systems due to hardware constraints [4]. With the advancement of deep neural networks (DNNs), learning-based approaches—especially convolutional neural networks (CNNs) [5]—have become mainstream in biomedical image denoising. CNNs perform well in extracting local features and reducing noise but struggle with capturing global dependencies due to limited receptive fields and pooling operations that discard spatial details. Additionally, most CNNs lack interpretability, a critical requirement in clinical applications where explainability is essential for decision support [6]. To overcome these limitations, recent work has explored attention-based models like Transformers, which offer better global feature modeling and can preserve structural integrity in complex medical images such as LDCT, MRI, or ultrasound. These models improve both performance and clinical relevance [7].

Transformer models have achieved remarkable results in computer vision and show potential to surpass CNNs. They excel at capturing global information and long - range feature interactions, enabling them to leverage richer data. As depicted in Figure 1, Transformers generate more effective features than CNNs. In biomedical image processing, where precision is crucial, Transformers can identify subtle patterns in complex images like CT scans and MRIs, aiding in tasks such as tumor detection and disease diagnosis. Their self - attention mechanisms also enhance the reliability of AI - assisted diagnostic tools by providing better visual interpretability [8].

CTFormer performs well in low-dose CT denoising by capturing long-range dependencies, but struggles with local detail and multi-scale structure preservation. To address this, we propose CTLformer with two key innovations. First, we introduce a multi-scale attention mechanism with local-global interaction, incorporating a multi-scale Token generation module within Token2Token to capture both fine (small-scale) and global (large-scale) features. In the self-attention module, an interaction unit distinguishes local and global attention ranges, alternately enhancing their weights to better preserve textures and structures. Second, we propose a dynamic attention control mechanism to handle varied noise distributions. A lightweight module (MLP or convolution) computes dynamic adjustment factors based on image noise or texture, reallocating attention accordingly. This mechanism emphasizes high-noise areas while preserving low-noise regions, improving both denoising and detail retention. Together, these enhancements significantly boost CTLformer's adaptability, robustness, and performance in real-world clinical applications.

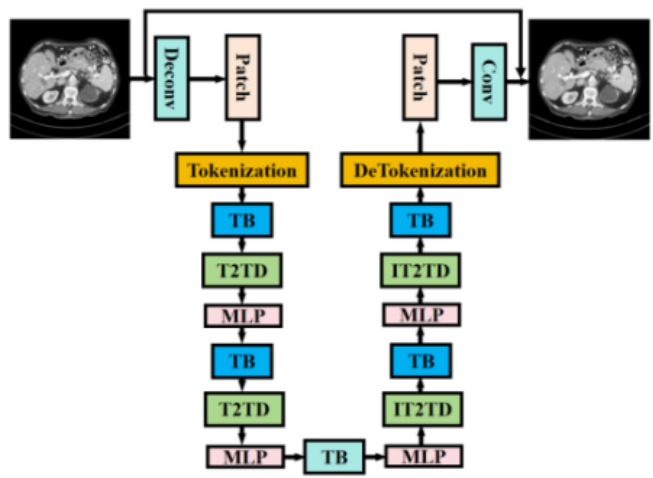

**Figure 1.** Architecture of our method

# 2. METHODOLOGY

*2.1. Multi-Scale Attention Mechanism with Local-Global Interaction*

To address the challenge of preserving fine details and multi-scale structures in low-dose CT images, we propose a multi-scale attention mechanism with local-global interaction[9]. This mechanism is implemented through two main components: the Token2Token mechanism and an enhanced self-attention module with interaction capabilities. The Token2Token mechanism is designed to capture fine details and global structures at different scales. The process begins with generating multi-scale tokens that represent both fine details (small-scale features) and global structures (large-scale features). This is done by applying convolution operations to the input CT image to extract multi-resolution feature maps. These feature maps are then used to generate tokens that capture information at different resolutions. The generated tokens are passed through the Token2Token mechanism, where each token represents either a fine detail (e.g., edges or textures) or a global feature (e.g., background or large structures). The interaction between these tokens allows for the exchange of information across different scales, ensuring that both fine details and global structures are preserved throughout the model. This multi-scale representation is crucial for maintaining image quality in low-dose CT images, where fine details are often lost.

This interaction module enables the model to focus on edge features and fine details when necessary, while also attending to the global context to preserve the overall image structure. By balancing these two aspects, the model can improve noise reduction and enhance the preservation of image quality in noisy low-dose CT scans.

*2.2. Adaptive Noise-Aware Attention Mechanism*

In addition to the multi-scale attention mechanism, we introduce a dynamic attention control mechanism to address the challenge of varying noise distributions across CT images. Low-dose CT images can exhibit different levels of noise, which can hinder the model's ability to preserve fine details. To adapt to these varying noise levels, we propose a dynamic weighting adjustment module integrated within the Transformer block of CTformer.

The dynamic weighting adjustment module is designed to adjust the attention distribution dynamically based on the noise characteristics and texture features of the input image. A lightweight network, a Fully connected layer, is used to generate dynamic attention adjustment factors. These factors are derived from the noise patterns of the input image and are used to reallocate the attention weights across different regions of the image.

This dynamic adjustment allows the model to focus more on high-noise regions, where noise suppression is more critical, while allocating less attention to low-noise areas where detail preservation is more important. The ability to adaptively adjust the attention ensures that the model performs well across a variety of noise distributions, making it more suitable for clinical applications where noise patterns can vary significantly.

To further enhance the model's noise reduction capabilities, we prioritize high-noise regions by increasing the attention on these areas. Using the dynamic attention adjustment factors, we detect noisy regions within the input image and reallocate attention weights accordingly. This ensures that high-noise areas receive more focus for suppression, while low-noise areas maintain fine details.

This approach not only improves noise reduction but also ensures that the model can preserve fine details in less noisy regions. The dynamic adjustment of attention weights allows the model to adapt to diverse noise environments, improving its robustness and effectiveness in real-world clinical scenarios.

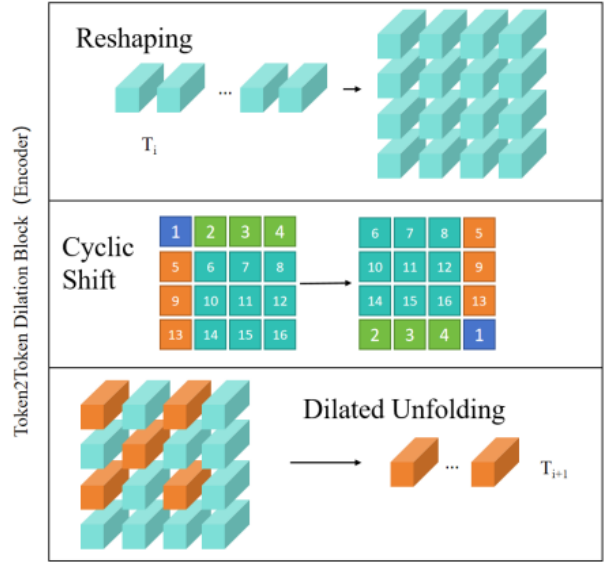

**Figure 2.** Architecture of Token2Token Block

## 3. Experiments and Analysis

### 3.1. Dataset

Model training and testing relied on the publicly available dataset from the 2016 National Institutes of Health AAPM Mayo Clinic LDCT Challenge [10]. The dataset comprises 2,378 low-dose (quarter-dose) and normal-dose (full-dose) CT images with a 3.0mm slice thickness, sourced from 10 anonymized patients. For model evaluation, cross-validation was implemented, allocating images from 9 patients to the training set and reserving those from 1 patient for testing. Moreover, as part of the data augmentation strategy, the original images underwent random rotations and flipping to enhance dataset diversity.

### 3.2. Evaluation Metrics

In biomedical image processing, especially in tasks like low-dose CT (LDCT) and MRI reconstruction, accurate and perceptually meaningful quality assessment is essential. To evaluate the performance of different models, we adopted three widely used quantitative metrics: Root Mean Square Error (RMSE), Peak Signal-to-Noise Ratio (PSNR), and Structural Similarity Index Measure (SSIM), and also compared the number of parameters across models. RMSE measures the average magnitude of pixel-wise errors between the reconstructed and reference images, with larger values indicating poorer reconstruction accuracy. PSNR, derived from Mean Squared Error (MSE), quantifies image fidelity by comparing the maximum possible signal to the noise level; higher values suggest less distortion. Unlike RMSE and PSNR, SSIM focuses on structural information, luminance, and contrast, offering a perceptual similarity score between 0 and 1, where a higher score indicates better structural preservation. Since biomedical images require high structural fidelity for clinical diagnosis, SSIM is particularly valuable in assessing the perceptual quality of reconstructed images. Additionally, model parameter comparisons help balance performance and computational efficiency for practical deployment [11].

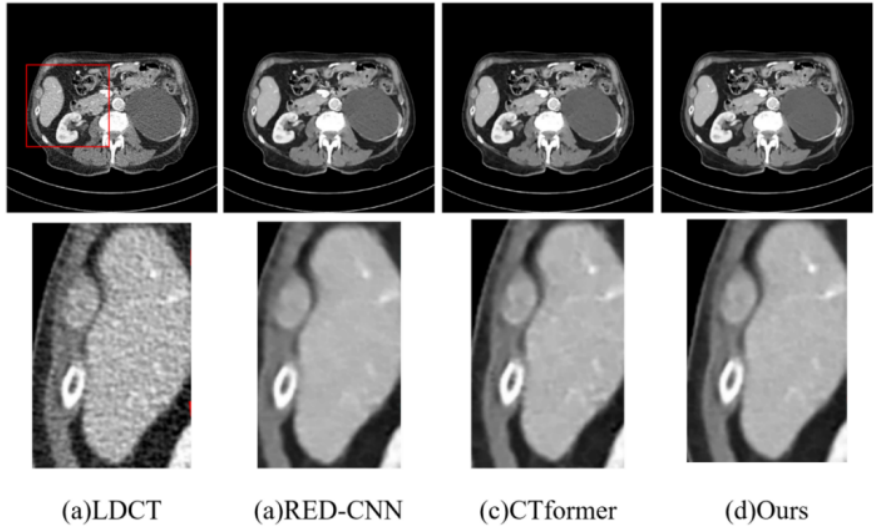

**Figure 3.** Denoising results using the proposed method on L506.

*3.3.Implementation Details*

The experiments are conducted on a platform running Windows 11, with a 13th Gen Intel(R) Core(TM) i9-13900HX, 2.20 GHz. The model is implemented using PyTorch 2.0.0 and CUDA 11.3. Training is performed on a GeForce RTX 4080 GPU platform with a batch size of 4. The initial learning rate is set to 0.0001.

*3.4.Experiment Result*

To comprehensively assess the denoising performance of the proposed CTLformer model in low-dose CT (LDCT) imaging, we selected slice 057 from patient L506 in the test set as a representative case, along with its corresponding region of interest (ROI) images. As shown in Figure 3 and detailed in Table 1, noise is mainly concentrated in the abdominal region, severely affecting the clarity of organ boundaries and tissue structures. Striped artifacts in the liver and spine regions further complicate clinical diagnosis. Among the compared methods, RED-CNN—a residual encoder-decoder convolutional neural network—effectively reduces high-frequency noise while preserving some texture details. However, due to its limited receptive field and reliance on local convolutions, it struggles to recover global anatomical structures. CTformer, built on a Transformer architecture, models long-range dependencies well but lacks convolutional layers, leading to over-smoothed textures and blurred soft-tissue details. In contrast, CTLformer integrates the strengths of both convolution and attention mechanisms, achieving a better balance between noise suppression and detail preservation. Quantitatively, it outperforms CTformer with a higher SSIM (0.9141 vs. 0.9120) and a lower RMSE (9.0133 vs. 9.0223), highlighting its superior reconstruction quality. In biomedical imaging, especially under low-dose conditions, maintaining fine structural details while suppressing noise is critical for diagnostic reliability, and CTLformer demonstrates strong potential in addressing this challenge.

Table 1. Comparison different algorithms on the L506.

| Method | SSIM | RMSE | params |
|---|---|---|---|
| LCDT | 0.8759 | 14.2416 | - |
| RED-CNN | 0.9077 | 10.1044 | 1.85M |
| WGAN-VGG | 0.9008 | 11.6370 | 34.07M |
| CTformer | 0.9120 | 9.0223 | 1.45M |
| CTLformer | 0.9141 | 9.0133 | 1.85M |

Table 1 presents quantitative comparisons of the proposed CTLformer against several existing models on test images from patient L506. The metrics evaluated include Structural Similarity Index (SSIM), Root Mean Square Error (RMSE), and the number of parameters (params). CTLformer[12] achieved the highest SSIM score of 0.9141, indicating superior perceptual similarity to ground truth images compared to all other tested models. Additionally, CTLformer recorded the lowest RMSE value of 9.0133, demonstrating its effectiveness in minimizing reconstruction errors, thereby confirming enhanced denoising performance.

When comparing CTLformer against RED-CNN, it is evident that while RED-CNN effectively reduces noise, its limited receptive field restricts its ability to restore global image structures, resulting in incomplete denoising. In contrast, CTformer, although proficient in capturing global context, suffers from excessive smoothing, leading to loss of fine details and blurred textures. CTLformer effectively overcomes these limitations by incorporating convolutional layers for local feature extraction and Transformer blocks for global context comprehension, thus preserving critical image details such as organ contours and textures while significantly reducing noise and artifacts. The parameter count further underscores the efficiency of CTLformer. With only 1.85 million parameters, CTLformer achieves superior results compared to WGAN-VGG and other methods, indicating an optimal balance between computational efficiency and performance.

## 4.Conclusion

In this paper, we introduce CTLformer, a novel low-dose CT (LDCT) denoising model that combines convolutional structures with transformer architecture. By incorporating a multi-scale attention mechanism and a dynamic attention control mechanism, CTLformer achieves significant improvements in denoising performance and model efficiency. The multi-scale attention mechanism captures both fine details and global structures through the Token2Token mechanism and self-attention interactions, enhancing important features while suppressing irrelevant noise. The dynamic attention control mechanism adapts attention based on noise distribution, focusing on high-noise areas and preserving details in low-noise regions, improving robustness. CTLformer also integrates convolutional layers for feature extraction and uses overlapping inference to reduce boundary artifacts. Experimental results show that CTlformer outperforms existing methods in both denoising and efficiency, providing an advanced solution for LDCT denoising with potential applications in medical image analysis.